\pdfoutput=1
\documentclass[10pt]{article}
\usepackage[a4paper, paperwidth=17.2cm, paperheight=25cm, inner=1.5cm, top=1.5cm, outer=1cm, bottom=1cm, includefoot]{geometry}
\usepackage{amsmath}
\usepackage{bm}
\usepackage[max2]{authblk}
\usepackage{amssymb}
\usepackage{amsmath}
\usepackage{graphicx}
\usepackage{enumitem}
\usepackage[ruled,lined]{algorithm2e}
\usepackage[numbers]{natbib}
\usepackage{amsthm}
\usepackage{appendix}
\usepackage{hyperref}
\usepackage{xcolor}
\usepackage{nameref}
\usepackage[nottoc]{tocbibind}
\usepackage{sectsty}

\newcommand{\sty}[1]{\boldsymbol{#1}}
\newcommand{\styy}[1]{\mathbb{#1}}

\let\epsilon\varepsilon
\let\mtheta\theta
\let\theta\vartheta

\let\rho\varrho

\let\phi\varphi
\let\Gamma\varGamma
\let\Delta\varDelta
\let\Theta\varTheta
\let\Lambda\varLambda
\let\Xi\varXi
\let\Pi\varPi
\let\Sigma\varSigma
\let\Upsilon\varUpsilon
\let\Phi\varPhi
\let\Psi\varPsi

\let\Omega\varOmega

\newcommand{\bmb}{\sty{b}}

\newcommand{\bme}{\sty{e}}

\newcommand{\bms}{\sty{s}}
\newcommand{\bmt}{\sty{t}}
\newcommand{\bmu}{\sty{u}}

\newcommand{\bmeps}{\sty{\epsilon}}

\newcommand{\bmsigma}{\sty{\sigma}}

\newcommand{\bbC}{\styy{C}}

\newcommand{\rmc}{\mathrm{c}}
\newcommand{\rmd}{\mathrm{d}}

\newcommand{\rmx}{\mathrm{x}}

\newcommand{\angstrom}{\text{\normalfont\AA}}

\title{AI-accelerated Materials Informatics Method for the Discovery of Ductile Alloys}
\author[1,2]{I. Novikov\thanks{i.novikov@skoltech.ru}}
\author[2]{O. Kovalyova\thanks{kovaleva.od@phystech.edu}}
\author[1]{A. Shapeev\thanks{a.shapeev@skoltech.ru}}
\author[1,3]{M. Hodapp\thanks{corresponding author}\thanks{maxludwig.hodapp@mcl.at}}
\affil[1]{Skolkovo Institute of Science and Technology (Skoltech), Center for Artificial Intelligence Technology, Moscow (RU)}
\affil[2]{Moscow Institute of Physics and Technology (MIPT), Moscow (RU)}
\affil[3]{Materials Center Leoben Forschung GmbH (MCL), Leoben (AT)}

\newcommand{\atom}{\ensuremath{r}}
\newcommand{\Atom}{\ensuremath{\sty{\atom}}}
\newcommand{\Atoms}{\ensuremath{\{\Atom_i\}}}

\subsectionfont{\normalfont\itshape}

\begin{document}

\maketitle

\begin{abstract}
In computational materials science, a common means for predicting macroscopic (e.g., mechanical) properties of an alloy is to define a model using combinations of descriptors that depend on some material properties (elastic constants, misfit volumes, etc.), representative for the macroscopic behavior.
The material properties are usually computed using special quasi-random structures (SQSs), in tandem with density functional theory (DFT).
However, DFT scales cubically with the number of atoms and is thus impractical for a screening over many alloy compositions.

Here, we present a novel methodology which combines modeling approaches and machine-learning interatomic potentials.
Machine-learning interatomic potentials are orders of magnitude faster than DFT, while achieving similar accuracy, allowing for a predictive and tractable high-throughput screening over the whole alloy space.
The proposed methodology is illustrated by predicting the room temperature ductility of the medium-entropy alloy Mo-Nb-Ta.
\end{abstract}


\tableofcontents

\section{Introduction}

In recent years, the class of random alloys, and in particular its subclass, the high-entropy alloys, has gained an exponentially increasing interest due to the promising opportunity to advance the current limits of mechanical properties (yield strength, fracture toughness, hardness, etc.) of classical alloys \citep{george_high_2020,ma_unusual_2020,curtin_progress_2022}.
This opportunity grounds on the unconstrained composition of a random alloy, possibly allowing for many elements to exist at (close-to) equi-atomic concentration, in comparison with classical alloys which usually contain one single principal element.
However, the vast amount of random alloys at different compositions can not be investigated with real experiments alone and requires a computational approach.
From the computational perspective, however, one of the main challenges is the number of possibilities of distributing elements within a configuration at a specific composition, since this number grows  superexponentially with the number of elements.

To overcome this curse of dimensionality, different methods have been developed over the past decades.
For example, average properties of on-lattice configurations can be efficiently simulated using the coherent potential approximation \citep[CPA,][]{yonezawa_coherent_1973,vitos_anisotropic_2001}.
However, the CPA does not account for relaxations which practically always influence the alloy properties (see, e.g., \citep{ikeda_ab_2019}).
To account for relaxations, the only possibility is to generate an ensemble of configurations, with random distributions of the elements according to a specific composition, relax them, and finally average over the ensemble.
Reasonably accurate average values can already be obtained with $\sim$\,30 configurations using sophisticated special quasi-random structures (SQSs) or variants thereof \citep{zunger_special_1990,yang_modeling_2016}.
However, even assuming a rather small number of 50 relaxation steps for each configuration yields a total number of single-point density functional theory (DFT) calculations of more than one thousand---for only one composition!
Clearly, this does not allow for a high-throughput screening of random alloy properties.

A tractable approach that allows for relaxing huge amounts of configurations are machine-learning interatomic potentials (MLIPs). Contrary to empirical interatomic potentials, they provably approximate DFT energies, forces, and stresses, with systematic errors, depending on the number of basis functions \citep{behler_generalized_2007,bartok_gaussian_2010,thompson_spectral_2015,shapeev_moment_2016,smith_ani-1:_2017,schutt_schnet_2017,pun2019-pinn,jinnouchi2019-kresse-on-the-fly,park2020-gnn,lysogorskiy2021-PACE}.
This has been recognized by the computational materials community and triggered ongoing developments of methodology for constructing MLIPs for random alloys \citep{li_complex_2020,byggmastar_modeling_2021,yin_atomistic_2021}.
In \citep{hodapp_machine-learning_2021}, we have have developed our own method for constructing Moment Tensor Potentials \citep[MTPs,][]{shapeev_moment_2016}, a class of MLIPs, for random alloys using an active learning algorithm \citep{podryabinkin_active_2017,gubaev_accelerating_2019} that selects $\sim$\,100 configurations that are representative for all neighborhoods appearing in all relaxations, out of a very large set of more than 10\,000 training candidate configurations.
We have then shown that the MTP, trained on those $\sim$\,100 configurations, was able to predict stacking fault energies of ternary random alloys, with differences up to a few percent with respect to very expensive reference DFT calculations.

In this work, we present one possible application of our method: an efficient parameterization of material models that predict mechanical properties of a random alloy.
More precisely, we parameterize the model for room temperature ductility of bcc random alloys from \citep{li_ductile_2020,mak_ductility_2021}, which is based on the ratio of the $K$-factors for dislocation emission and cleavage, for the Mo-Nb-Ta medium-entropy alloy.
This model was shown to be in good qualitative agreement with real experiments, and depends on the stacking fault energies on \{112\} planes, the \{110\} surface energies, and the elastic constants.%
\footnote{In principle, the model also depends on other orientations, but it was shown in \citep{mak_ductility_2021} that the \{112\} stacking fault plane and the \{110\} surface plane are the most important ones for Mo-Nb-Ta}
Our approach here is to construct a separate MTP for each of those material properties using the algorithm from \citep{hodapp_machine-learning_2021}.
That is, we first validate the trained MTPs for the stacking fault energies on \{110\} planes and the \{110\} surface energies---because we have reference DFT values for those orientations---by showing that the MTPs predict the DFT energies with only a few percent deviation.
We then make predictions for room temperature ductility of Mo-Nb-Ta over the entire composition space.

On a side track, we introduce and validate an average-atom MTP, an \emph{"alchemical MTP"}, which has, as will be shown, material properties close to the random alloy MTP.
This allows for further speeding up the computation of the material properties for many alloy compositions, and we anticipate that it will prove useful in future applications where a desired average property can not be easily obtained by averaging over true random configurations, such as the average line tension of a random alloy.

\subsection{Ductility Model}

The ductility model we employ here is based on the postulate that ductility is controlled by the fracture behavior at an atomistically sharp crack in a pre-cracked material.
That is, a material is considered to be ductile if, upon mode I loading, the crack front starts blunting, due to dislocation emission, rather than cleaving.
This implies that ductility can be described as a competition between dislocation emission and crack propagation, and the critical stress state upon one of them occurs is defined by the stress intensity factors $K_{\rm Ie}$ and $K_{\rm Ic}$, respectively.

To compute $K_{\rm Ie}$ and $K_{\rm Ic}$, we assume linear elastic fracture mechanics such that this competition between dislocation emission and crack propagation can be expressed via the ductility index 
\begin{equation}\label{eq:ductility_index}
    D = \frac{K_{\rm Ie}(\gamma_{\rm sf},\bbC)}{K_{\rm Ic}(\gamma_{\rm surf},\bbC)},
\end{equation}
where both $K$-factors depend on the stacking fault energy $\gamma_{\rm sf}$, the surface energy $\gamma_{\rm surf}$, and the elastic constants $\mathbb{C}$ at 0\,K.
For further details on how to derive \eqref{eq:ductility_index} we refer to Section ``\nameref{sec:linear_elastic_fracture_mechanics}''.

As such, $D$ represents a \emph{qualitative measure} for comparing ductility trends between different materials.
In order to make \emph{quantitative predictions} further requires the introduction of a threshold below which a material is assumed to be ductile.
Such a threshold must be calibrated to real experiments (cf. \citep{mak_ductility_2021}) and, therefore, allows to \emph{implicitly} include finite temperature effects in our model.
We will return to this question later on when we will make actual predictions for Mo-Nb-Ta.

\subsection{Active Learning Algorithm for Computing the Ductility Index}

Our main goal is now to compute $D$ for any random alloy from the Mo-Nb-Ta family.
To that end, we will construct three MTPs, one for each property, that is, for the stacking fault energy, for the surface energy, and for the elastic constants.

\begin{figure}[t]
    \centering
    \includegraphics[width=0.99\textwidth]{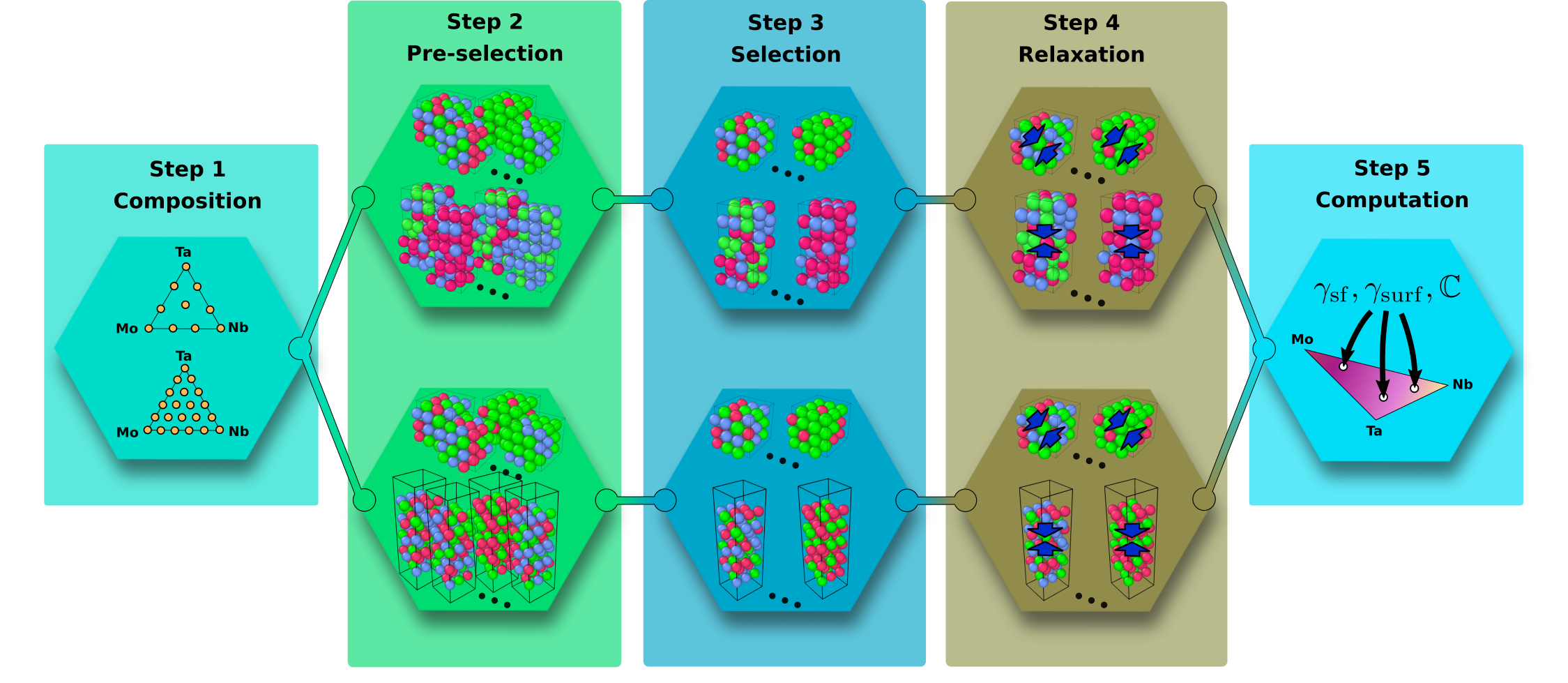}
    \caption{%
    Schematic illustration of the active learning algorithm for computing the ductility index.
    The core part, \textbf{Steps\;2--4}, consists of training three MTPs, one for computing the stacking fault energies (top), and one for computing the surface energies (bottom).
    The MTP for computing the elastic constants is constructed by adding additional strained bulk configurations (cf. Figure \ref{fig:cfgs} (b)) to the training set of the stacking fault MTP.} 
    \label{fig:algo}
\end{figure}

Our strategy to construct those three MTPs is as follows:
we first construct the two MTPs for computing the stacking fault and surface energies that are actively trained on DFT training sets containing configurations with either stacking faults or free surfaces, respectively.
Both training sets also contain bulk configurations.
The active learning algorithm for constructing those two MTPs is illustrated in Figure \ref{fig:algo} and described below.

\begin{itemize}
    \item {\bf Step 1.}
    We first create the composition domain in which we seek to construct the set of training candidate configurations.
    In the figure, a homogeneous grid is used, covering the entire composition space.
    \item {\bf Step 2.} We create two huge sets of more than 10\,000 training candidates, with different concentrations of Mo, Nb, and Ta, corresponding to the composition domain chosen in {\bf Step 1}, one for the stacking fault MTP, and one for the free surface MTP. We initialize both MTPs by training them on a small subset of these training candidate sets containing $\sim$\,100 configurations and initialize the active sets.
    \item {\bf Step 3.} We then select the most remaining representative configurations from the training candidate sets using active learning described in Section ``Materials and Methods'' below, conduct DFT calculations for these configurations, and add them to the initial training sets. We re-train the MTPs on these updated sets and re-initialize the active sets.
    \item{\bf Step 4.} We consecutively relax all the configurations in each training set with active learning ``switched on'', update the training sets, and obtain the final MTPs for calculating the stacking fault and free surface energies.
    \item{\bf Step 5.} In the final step, we calculate the stacking fault and free surface energies for all desired concentrations of Mo, Nb, and Ta.
\end{itemize}

For calculating the elastic constants, we update the MTP fitted to the configurations with stacking faults. That is, we create configurations with normal and shear deformations for different concentrations of Mo, Nb, and Ta, select some of them using the active set created in {\bf Step 3} of the algorithm for the stacking fault calculations, conduct DFT calculations for them, and re-train the MTP. With this re-trained MTP we calculate the elastic constants.

To compute the \emph{average} material properties, we propose two methods, (i) by directly computing them using random configurations with sufficiently large supercells of $>$\,1\,000 atoms, and (ii) by using an \emph{average-atom MTPs}.
More precisely, for method (ii), we construct three average-atom MTPs for a specific alloy composition, one for each property, by averaging the descriptors of the corresponding MTP over all possible neighborhoods for the specific alloy composition.
This yields three effectively \emph{single-component MTPs} that are supposed to possess material properties close to the average material properties of the random alloy, thus avoiding the need for large supercells.
More details on how we construct the average-atom MTPs are given in the ``Methods'' section.

\section{Results and Discussion}

\subsection{Detailed Description of the Active Learning Algorithm}

In the following, we present a detailed description of the active learning algorithm to construct the three MTPs that we use to predict the ductility indices for the Mo-Nb-Ta ternary random alloys ({\bf Steps\;1--5} from Figure \ref{fig:algo}).
We remark that, for the stacking fault energies, we first present the results for the stacking faults on \{110\} fault planes, and compare those with existing values from the literature \cite{zhou_misfit-energy-increasing_2004, hu_screening_2021}.
The results for the \{112\} fault planes, required for predicting the ductility indices, are presented afterwards. 

We start by creating the reference configuration for a bcc lattice. In this reference configuration, we then introduce \{110\} stacking faults and surface energies.
We then construct the composition domain according to {\bf Step\;1}, more precisely the compositions for which we create the configurations for the set of training candidates.
We choose the lower grid from Figure \ref{fig:algo} containing 21 points, and create three types of configurations: 21 bulk configurations of 54 atoms, 21 stacking fault configurations of 72 atoms, and 21 free surface configurations of 36 atoms.

Next, in {\bf Step\;2}, we randomly distribute the atoms of different types inside the supercell according to the desired compositions.
In total, we create 75 configurations, i.e., four random distributions of atoms for each of the 18 binary and ternary compositions, and three more configurations with pure Mo, Nb, and Ta. In addition, we apply random displacements to each of these binary and ternary configurations one time, and do the same for pure Mo, Nb, and Ta configurations four times. These random displacements are drawn from a normal distribution with standard deviation 0.0033$a_0$, where $a_0$ is the lattice constant for the corresponding composition. We then combine the bulk and the stacking fault configurations in the first initial training set, and the bulk configurations and the free surface configurations in the second initial training set.
Each of these initial sets contains 168 configurations. We then conduct the DFT calculations, and train the MTPs on these two initial training sets. 

We then create the training candidates to be selected and added to the initial training sets. To that end, we take 100 random distributions of atoms for the 18 binaries and ternaries, and randomly displace atoms in each of these configurations as described above. We also apply 100 random displacements to the single-component configurations. Finally, we compress and extend these configurations in order to cover the entire range of lattice constants. This procedure leads to 12600 candidate bulk configurations, stacking fault configurations, and free surface configurations. As for the initial sets, we combine the bulk configurations, first, with the stacking fault configurations and, second, with the free surface configurations, and we thus obtain two candidate sets of 25200 configurations each. In {\bf Step\;3}, we then use the active learning algorithm described in Section ``Materials and Methods'' below to select the extrapolative configurations from the candidate sets, run DFT calculations on them, add them to the initial training sets, and re-fit the MTPs.

In {\bf Step\;4} we relax all the configurations from the two training sets created in {\bf Step\;3}. We run the active learning algorithm during the relaxation and, thus, we select and add new configurations to the training sets, and re-train the MTPs. The relaxation terminates when the maximum force on all atoms is less than 1 meV/$\angstrom$. The resulting training sets for the MTPs used to calculate the stacking fault energy and the surface energy contain 228 and 241 configurations, respectively. The training errors are reported in Table \ref{Tab:Train_error_SF_SE_elastic}. Here, we use MTPs of level 16, since it was shown in \citep{hodapp_machine-learning_2021} that this MTP level is sufficient to accurately predict stacking fault energies of the Mo-Nb-Ta alloy. We approximate MTP energies and forces to the DFT ones during the MTP fitting. From the table we conclude that both MTPs for the stacking fault energy and the surface energy were accurately fitted, as all the energy errors are smaller than 2 meV/atom and the force errors are smaller than 40 meV/$\angstrom$ whereas the typical energy and force MAEs obtained after MLIPs fitting are about 2 meV/atom and 100 meV/$\angstrom$, respectively (see, e.g., \cite{zuo2020performance}).

For calculating the ductility indices, we also need to accurately predict the elastic constants. To do so, we enrich the training set containing the stacking fault configurations. First, as in {\bf Step 1}, we additionally create 45 bulk configurations of 72 atoms with different compositions. Next, we extend, compress, or shear, the configurations by 2 \%, i.e., we create 12 additional configurations with extensions/compressions along the $xx$, $yy$, $zz$, $yz$, $xz$, $xy$ directions (i.e., strained configurations) for each of the composition. Finally, we apply random displacements to each of these configurations five times. As a result, we have 2925 candidates to be added to the stacking fault training set. After that, like in {\bf Step 2}, we select the configurations from this set of the candidates, run DFT calculations for the selected configurations, add these configurations to the stacking fault set, and, finally, we re-train the MTP. We note that here we fit not only MTP energies and forces to the DFT ones, but also stresses as we predict elastic constants with this MTP. The resulting training set with the stacking faults, shear, and bulk deformations, contains 240 configurations. The training errors are also given in Table \ref{Tab:Train_error_SF_SE_elastic}. From the table we conclude that the MTP was trained with reasonable accuracy. We also emphasize that we totally have less than 1000 configurations in all the three training sets and, therefore, we need less than 1000 expensive single-point DFT calculations for fitting the MTPs used to calculate the properties of interest for the ternary alloy. Thus, the algorithm proposed here allows us to optimize the number of DFT calculations and to avoid adding an excessive amount of (geometrically similar) configurations to the training sets.

\begin{table}[h!]
\centering
\caption{\label{Tab:Train_error_SF_SE_elastic} The training mean average error (MAE) and root-mean square error (RMSE) for the MTPs fitted to the training sets including the stacking fault configurations, the ones with free surfaces, and the strained configurations.}
\begin{tabular}{|c|c|c|c|c|c|c|} \hline
\small{Training set} & \small{energy} \footnotesize{MAE,} & \small{energy} \footnotesize{RMSE,} & \small{force} \footnotesize{MAE,} & \small{force} \footnotesize{RMSE,} & \small{stress} \footnotesize{MAE,} & \small{stress} \footnotesize{RMSE,} \\ 
 & \small{meV/atom} & \small{meV/atom} & \small{meV/$\angstrom$} & \small{meV/$\angstrom$ (\%)} & \small{GPa} & \small{GPa (\%)} \\ \hline \hline
\small{Stacking fault} & \small{1.4} & \small{1.7} & \small{19} & \small{39 (9\%)} & - & - \\ \hline
\small{Free surface} & \small{1.4} & \small{1.8} & \small{18} & \small{35 (9\%)} & - & - \\ \hline
\small{Strained} & \small{1.0} & \small{1.3} & \small{13} & \small{28 (7\%)} & \small{0.044} & \small{0.165 (2 \%)} \\ \hline
\end{tabular}
\end{table}

Finally, in {\bf Step 5}, we calculate the properties to predict the ductility indices.
For computing the stacking fault and surface energies, we use one very large configuration with 10\,368 atoms, and for computing the elastic constants one of 2\,000 atoms.

To construct the average-atom MTPs for the three properties at a desired composition, we take the corresponding MTP from above and average its descriptors, as described in Section ``Average-atom MTPs''.
The stacking fault energies, surface energies, and elastic constants, are then computed as with any other single-component MTP.

\subsection{\texorpdfstring{$\{110\}$}{} Surface Energies}

\begin{figure}[h!]
\centering
\includegraphics[width=0.5\textwidth]{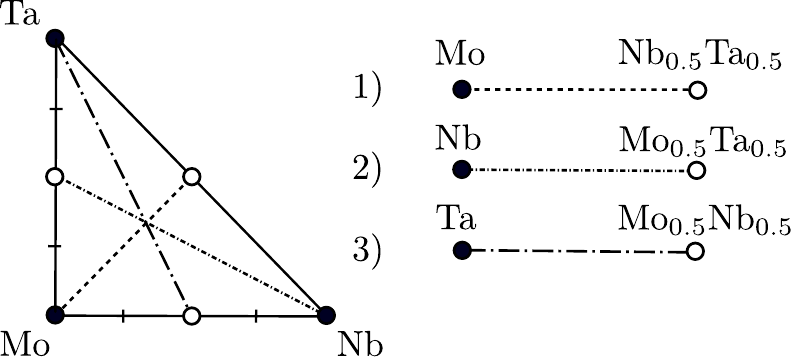}
\caption{The three curves through the Mo-Nb-Ta ternary diagram along which we compute the stacking fault and surface energies.}
\label{Fig:ternary} 
\end{figure}

First, we present the results for the surface energy calculations. We calculate the surface energies for various compositions along the three curves through the ternary diagram shown in Figure \ref{Fig:ternary}: Mo$\to$Nb$_{0.5}$Ta$_{0.5}$, Nb$\to$Mo$_{0.5}$Ta$_{0.5}$, and Ta$\to$Mo$_{0.5}$Nb$_{0.5}$.

\begin{figure}[t]
\centering
\includegraphics[width=2.7in, height=2.2in, keepaspectratio=false]{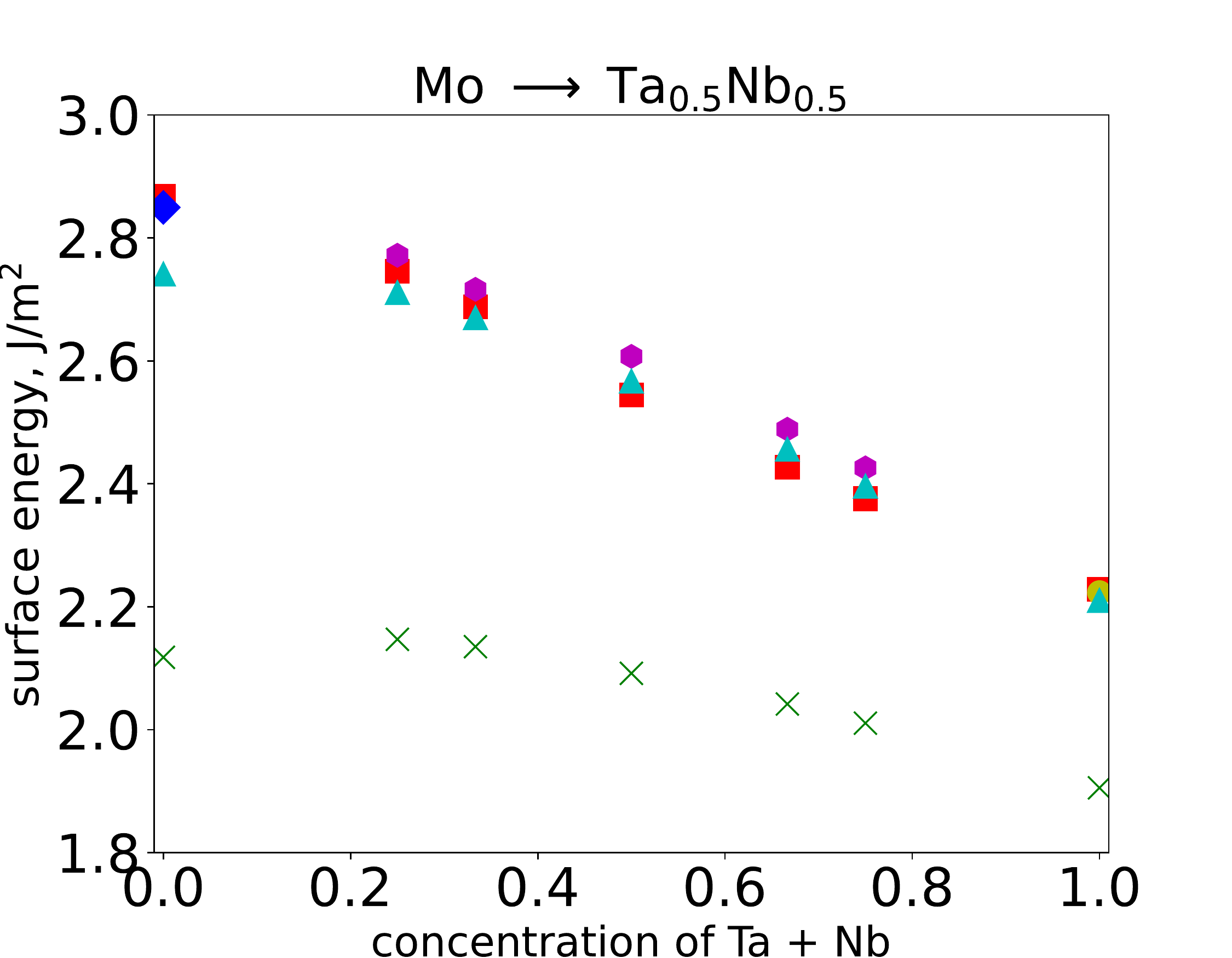}
\includegraphics[width=2.7in, height=2.2in, keepaspectratio=false]{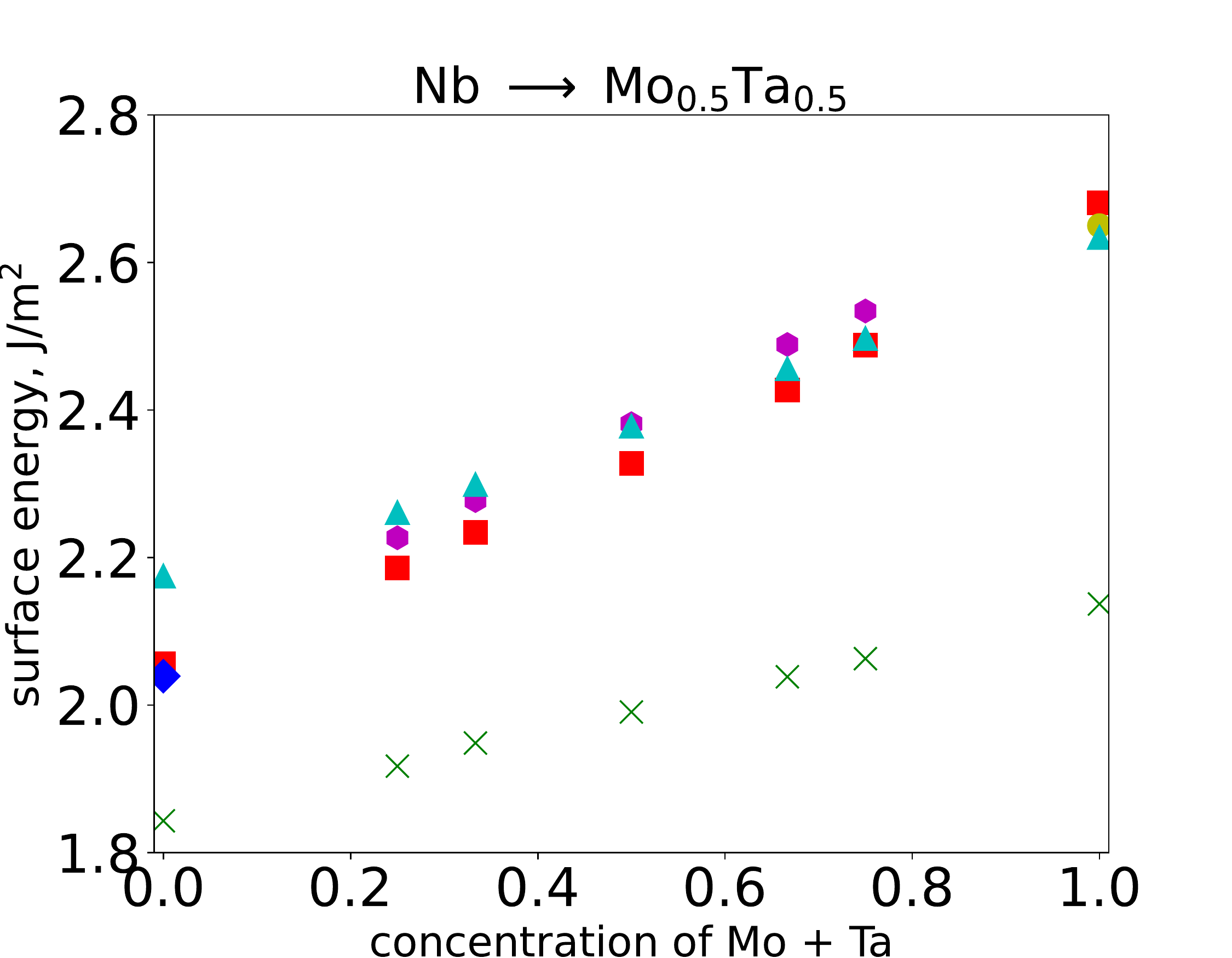}
\includegraphics[width=2.7in, height=2.2in, keepaspectratio=false]{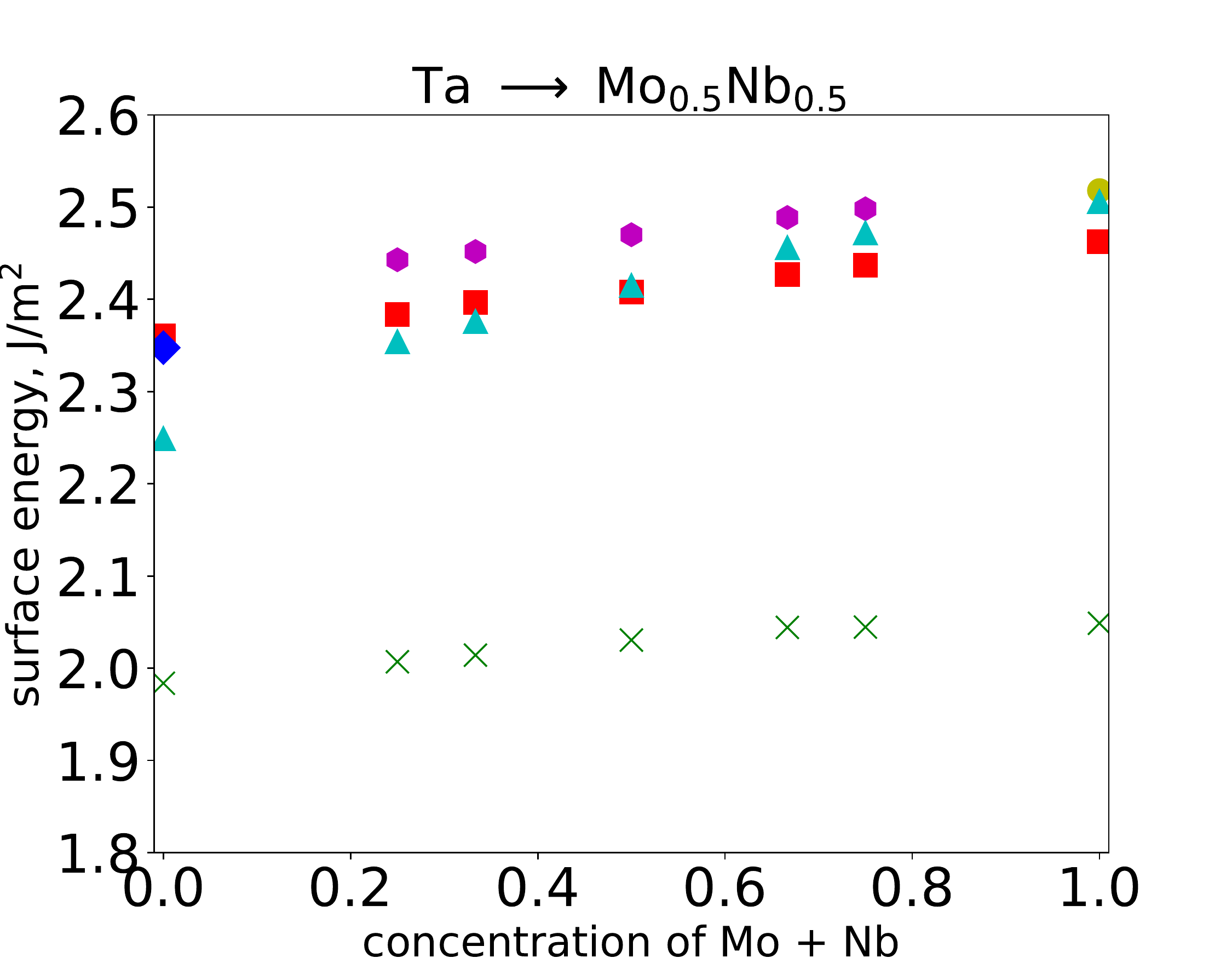}
\includegraphics[width=2.4in, height=2.0in, keepaspectratio=false]{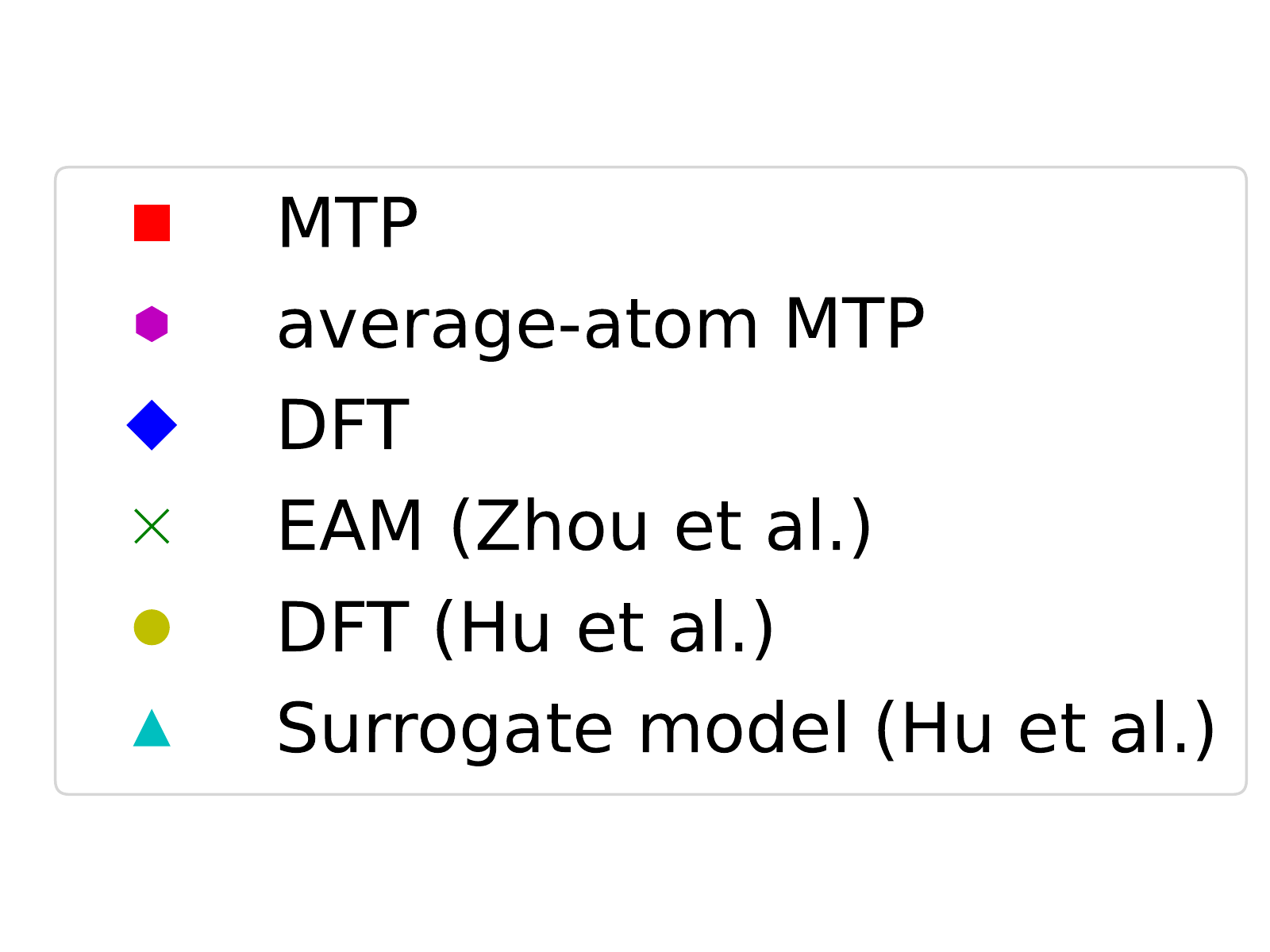}
\caption{\{110\} surface energies calculated with different models for various compositions of the Mo-Nb-Ta alloy. The random alloy MTP and average MTP surface energies are close to the ones computed with the surrogate model. The EAM potential significantly underestimates the surface energies.}
\label{Fig:Surf}
\end{figure}

We calculate the surface energies with the MTP fitted to the configurations with free surfaces obtained with the VASP \cite{kresse1993ab,kresse1994ab,kresse1996efficiency,kresse1996efficient} package (see the computational details in the ``Methods'' section). We compare the MTP surface energies with the ones calculated with the EAM potential from \cite{zhou_misfit-energy-increasing_2004} and with the surrogate model of \citet{hu_screening_2021} that has been fitted to an extensive set of DFT calculations.
\citet{hu_screening_2021} used a similar DFT setup compared to ours and showed that their surrogate model gives relative errors of only a few percent with respect to a test set containing up to quarternary compositions, thus, providing a reliable reference for the validation of our MTPs. In addition, we also calculate the surface energies for pure Mo, Nb, and Ta, with DFT. The results are shown in Fig. \ref{Fig:Surf}. From the figure we conclude that the MTP surface energies are close to the ones obtained with the surrogate model and DFT for different concentrations of Mo, Nb, and Ta, whereas the EAM potential significantly underestimates the surface energies. This result is expected as the EAM potential was not parameterized for predicting the surface energies.

The average-atom MTP is in excellent agreement with the random alloy MTP, with minor differences of a few percent for all considered compositions.
It is expected that local relaxations play a minor role here (since interactions with the free surface dominate), so the results in Figure \ref{Fig:Surf} imply that the MTP descriptors are indeed strongly uncorrelated.

\subsection{\texorpdfstring{$\frac{1}{4}\langle111\rangle\{110\}$}{} Stacking Fault Energies}

Next, we present the results for the stacking fault calculations.
While we have already computed them with MTPs in our previous publication \citep{hodapp_machine-learning_2021}, we have not constructed those MTPs using a \emph{uniform initialization} of the training set in \textbf{Step\;2}, which is essential as pointed out in \citep{hodapp_machine-learning_2021}.%
\footnote{otherwise the active learning algorithm may erroneously undersample certain compositions not considered in the initial training set}
So, the results shown here are new and provide an additional validation.

\begin{figure}[t]
\centering
\includegraphics[width=2.7in, height=2.2in, keepaspectratio=false]{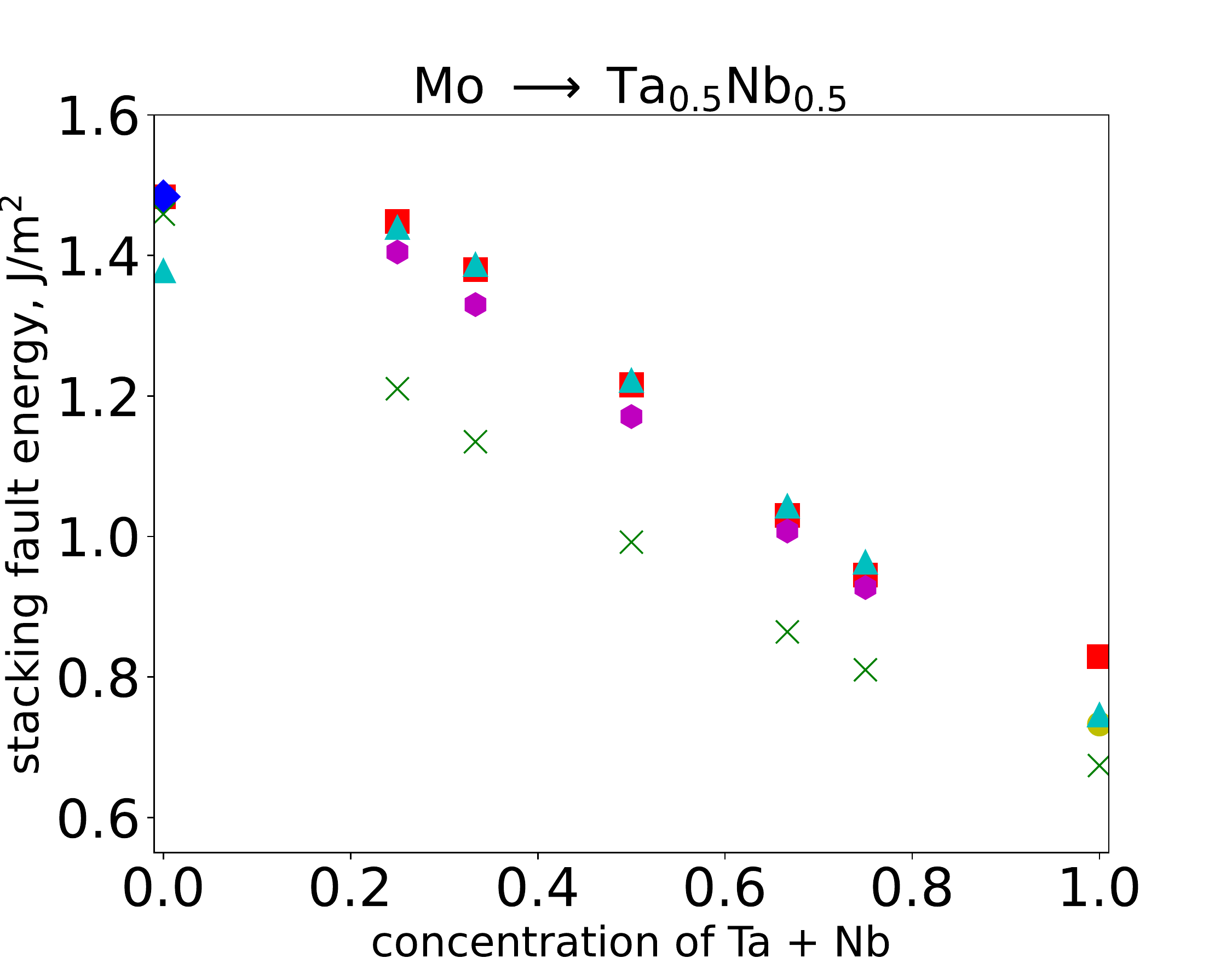}
\includegraphics[width=2.7in, height=2.2in, keepaspectratio=false]{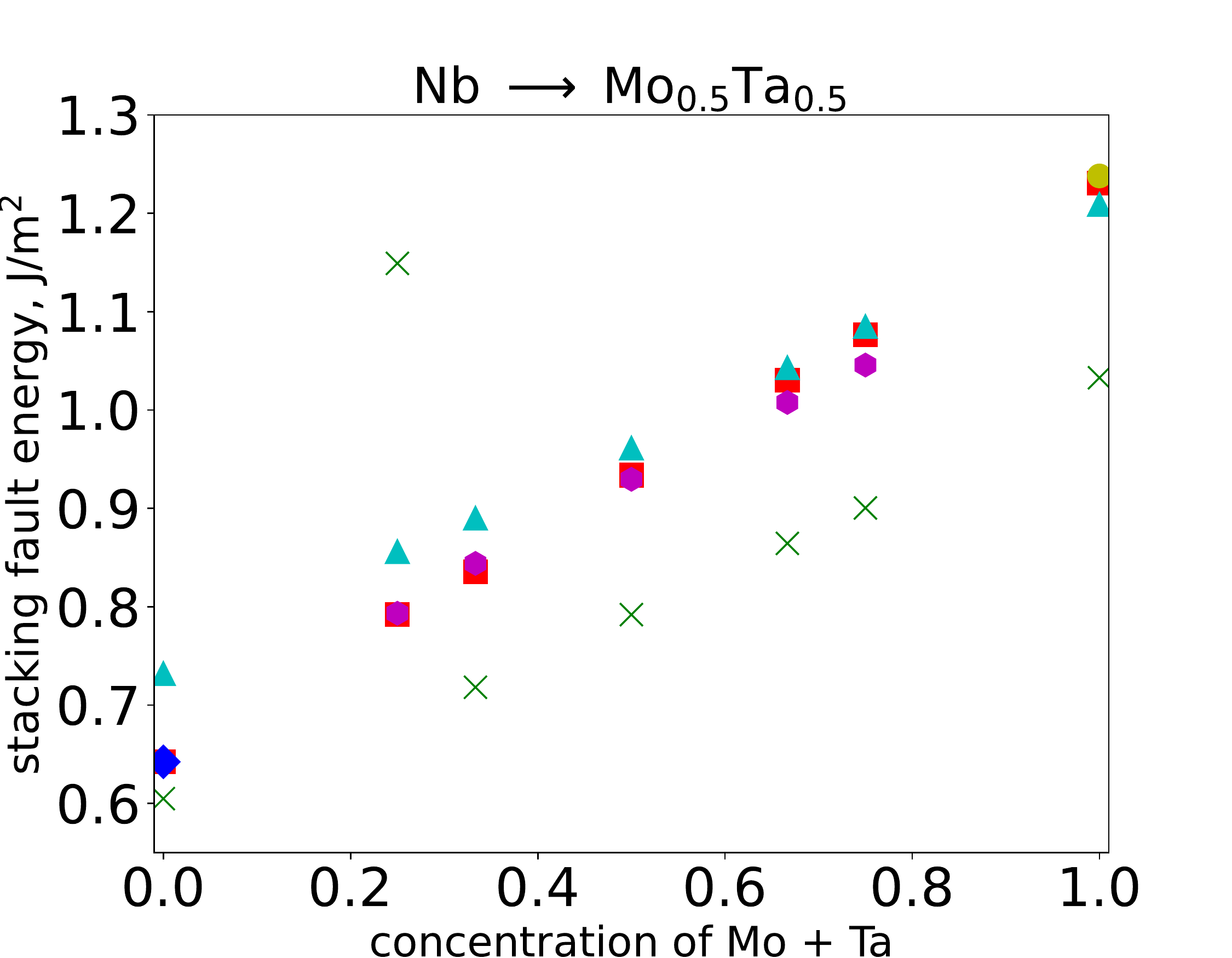}
\includegraphics[width=2.7in, height=2.2in, keepaspectratio=false]{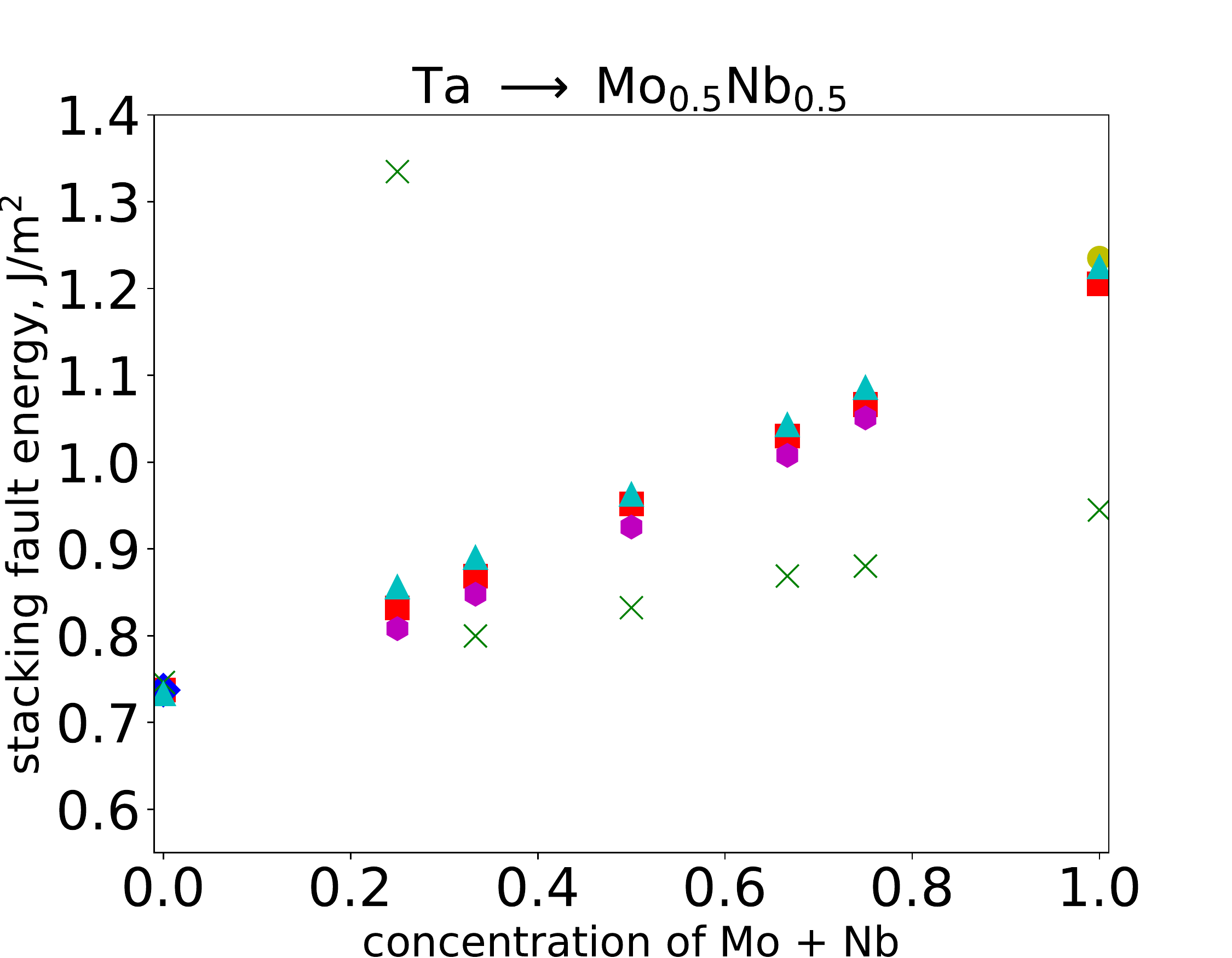}
\includegraphics[width=2.4in, height=2.0in, keepaspectratio=false]{Legend.pdf}
\caption{$\frac{1}{4}\langle111\rangle\{110\}$ stacking fault energies calculated with different models for various compositions of the Mo-Nb-Ta alloy. The random alloy MTP, the average-atom MTP, and the surrogate model, give close results, while the EAM potential mostly underestimates the stacking fault energies.}
\label{Fig:SFE}
\end{figure}

We analyze the same compositions of the Mo-Nb-Ta alloy and compare the stacking fault energies obtained with the models as in the previous subsection, except for the MTP: here we use the one trained on the stacking fault training set. The results are given in Fig. \ref{Fig:SFE}. From the figure we conclude that, as for the surface energies, the MTP stacking fault energies are close to the ones obtained with the surrogate model and DFT for different concentrations of Mo, Nb, and Ta, whereas the EAM potential mostly underestimates the surface energies. Again, this result is expected as the EAM potential was not parameterized for predicting the stacking fault energies.
As for the surface energies, the agreement between the average-atom MTP and the random alloy MTP is very good.

\subsection{Elastic Constants}

We now present the results of the elastic constants calculations. The results are given in Table \ref{Tab:Elastic}. We first compare the elastic constants obtained for pure Mo, Nb, and Ta with the EAM potential \cite{zhou_misfit-energy-increasing_2004}, with DFT, and experimental results \cite{bolef1961elastic,carroll1965elastic,featherston1963elastic} (see the lines 2--10 in the table). DFT typically underestimates the C$_{44}$ elastic constants (see, e.g., \cite{kovci2008-elasticity}). On the other hand, all the elastic constants calculated with the EAM potential are in a good agreement with the experimental ones. Therefore, the elastic constants obtained with the MTP fitted to DFT can differ from the ones calculated with the EAM potential (see the lines 11--32 in Table \ref{Tab:Elastic}). Indeed, from Table \ref{Tab:Elastic} we see that the C$_{11}$'s and C$_{12}$'s obtained with the MTP (average MTP) and the EAM potential are close (the RMSEs are 21 (23) GPa and 9 (14) GPa for the C$_{11}$'s and C$_{12}$'s, respectively), but, for C$_{44}$, the differences are rather big, between 50 and 100\%, namely, RMSE is 31 (34) GPa. For all the considered compositions the MTP underestimates C$_{44}$.

As for the stacking fault and surface energies, the agreement between the random alloy MTP and the average-atom MTP is good.
Relative differences can be up to $\sim$\,10\%, in particular for C$_{44}$, primarily due to the local relaxations.
Those differences are, however, tolerable, within the range of previously reported values for the Fe-Ni-Cr system \citep{varvenne_average-atom_2016}.

\begin{table}[h!]
\centering
\caption{\label{Tab:Elastic} Comparison of the C$_{11}$, C$_{12}$, and C$_{44}$ elastic constants calculated with different models and methods (the values in parentheses are the elastic constants predicted by the average-atom MTP, which are in good agreement with the random alloy MTP). DFT underestimates the C$_{44}$ elastic constants for pure elements, whereas the EAM potential and the experimental elastic constants are close to each other. The C$_{11}$'s and C$_{12}$'s calculated with the MTPs and the EAM potential for alloys are close to each other, but the difference between the C$_{44}$'s is rather big.}
\begin{tabular}{|c|c|c|c|} \hline
Element or Alloy/Method or Model & C$_{11}$, GPa & C$_{12}$, GPa & C$_{44}$, GPa \\ \hline \hline 
Mo/DFT & 490 & 161 & 96 \\ 
Mo/EAM & 457 & 167 & 113 \\ 
Mo/Experiment & 450 & 173 & 125 \\ \hline
Nb/DFT & 225 & 180 & 13 \\ 
Nb/EAM & 262 & 125 & 36 \\ 
Nb/Experiment & 253 & 133 & 31 \\ \hline
Ta/DFT & 273 & 164 & 72 \\ 
Ta/EAM & 263 & 158 & 82 \\ 
Ta/Experiment & 266 & 158 & 87 \\ \hline
MoNbTa$_2$/MTP (average-atom MTP) & 281 (281) & 166 (171) & 47 (42) \\
MoNbTa$_2$/EAM & 287 & 160 & 81 \\ \hline
MoNbTa$_4$/MTP (average-atom MTP) & 250 (251) & 164 (169) & 44 (39) \\
MoNbTa$_4$/EAM & 274 & 160 & 82 \\ \hline
MoNbTa$_6$/MTP (average-atom MTP) & 234 (236) & 163 (169) & 42 (38) \\
MoNbTa$_6$/EAM & 269 & 159 & 82 \\ \hline
Mo$_3$Nb$_2$Ta$_3$/MTP (average-atom MTP) & 323 (324) & 174 (177) & 54 (49) \\
Mo$_3$Nb$_2$Ta$_3$/EAM & 311 & 164 & 84 \\ \hline
MoNbTa/MTP (average MTP) & 310 (310) & 169 (172) & 51 (47) \\
MoNbTa/EAM & 304 & 160 & 79 \\ \hline
MoNb$_2$Ta/MTP (average-atom MTP) & 284 (285) & 159 (162) & 44 (41) \\
MoNb$_2$Ta/EAM & 292 & 152 & 68 \\ \hline
Mo$_2$Nb$_3$Ta$_3$/MTP (average-atom MTP) & 283 (283) & 163 (167) & 46 (41) \\
Mo$_2$Nb$_3$Ta$_3$/EAM & 289 & 156 & 74 \\ \hline
Mo$_2$NbTa/MTP (average-atom MTP) & 362 (366) & 179 (184) & 62 (60) \\
Mo$_2$NbTa/EAM & 339 & 167 & 88 \\ \hline
Mo$_4$NbTa/MTP (average-atom MTP) & 408 (415) & 183 (190) & 74 (74) \\
Mo$_4$NbTa/EAM & 379 & 170 & 96 \\ \hline
Mo$_6$NbTa/MTP (average-atom MTP) & 427 (433) & 179 (187) & 79 (81) \\
Mo$_6$NbTa/EAM & 399 & 171 & 101 \\ \hline
Mo$_3$Nb$_3$Ta$_2$/MTP (average-atom MTP) & 323 (324) & 169 (173) & 51 (50) \\
Mo$_3$Nb$_3$Ta$_2$/EAM & 313 & 160 & 78 \\ \hline \hline
\end{tabular}
\end{table}

\subsection{Ductile Compositions of Mo-Nb-Ta}

Motivated by the results from the previous sections, we now make predictions for room temperature ductility of random alloys.
Following \citep{mak_ductility_2021}, the most important orientation for alloys from the Mo-Nb-Ta family is \{110\}/\{112\} (crack plane/slip plane).
This is due to the higher \{100\} surface energies, in comparison with the \{110\} surface energies (cf., \citep{mak_ductility_2021}, Table 1).
We thus anticipate that it suffices to predict ductility using the ductility indices for \{110\}/\{112\}.

\begin{table}[h!]
\centering
\caption{\label{Tab:Train_error_planes} The training mean average error (MAE) and root-mean square error (RMSE) for the MTPs fitted to the training sets including the stacking fault configurations for the \{110\} and \{112\} planes, sizes of the training sets.} 
\begin{tabular}{|c|c|c|c|c|c|c|} \hline
\small{plane}/\small{size} & \small{energy} \footnotesize{MAE,} & \small{energy} \footnotesize{RMSE,} & \small{force} \footnotesize{MAE,} & \small{force} \footnotesize{RMSE,} & \small{stress} \footnotesize{MAE,} & \small{stress} \footnotesize{RMSE,} \\ 
 & \small{meV/atom} & \small{meV/atom} & \small{meV/$\angstrom$} & \small{meV/$\angstrom$ (\%)} & \small{GPa} & \small{GPa (\%)} \\ \hline \hline
\small{\{110\}}/\small{228} & \small{1.4} & \small{1.7} & \small{19} & \small{39 (9 \%)} & \small{-} & \small{-} \\ \hline
\small{\{112\}}/\small{147} & \small{1.8} & \small{2.2} & \small{24} & \small{51 (12\%)} & \small{0.26} & \small{0.91 (10 \%)} \\ \hline
\end{tabular}
\end{table}

Therefore, we have computed a fresh MTP to predict the stacking fault energies for the \{112\} plane.
To construct this MTP, we have essentially followed the same procedure as for the \{110\} plane, so we do not reiterate the details here.
The sizes of the training sets that include the stacking fault configurations for the \{110\} and \{112\} planes, and the corresponding training errors, are given in Table \ref{Tab:Train_error_planes}. From the table we conclude that both, the MTP for the \{110\} plane, and the MTP for the \{112\} plane, were trained with high accuracy. The energy and force errors for the \{112\} plane are a bit greater than for the \{110\} plane as we also fitted MTP stresses to DFT stresses for the \{112\} plane. Due to the same reason, the size of the training set for the \{112\} plane is smaller than the one for the \{110\} plane. We also note, that the stress errors for the \{112\} plane are small enough and, thus, once we have an accurate DFT method for calculating the C$_{44}$ elastic constants we will be able to automatically create machine-learning interatomic potentials for accurate predicting the ductility indices. Here, for computing the ductility index \eqref{eq:d_index}, we use the elastic constants obtained with the EAM potential from \cite{zhou_misfit-energy-increasing_2004},  since DFT does not predict the shear modulus of Nb, as shown in the previous section.
An MTP fitted to the elastic constants obtained from DFT is thus prone to errors, unlike the EAM potential that has been fitted to the experimental elastic constants.

To predict ductile compositions, we define a critical ductility index $D_\rmc$ below which the alloy is assumed to be ductile, and above which the alloy is assumed to be brittle.
Here, we adopt $D_\rmc$\,=\,$1.26$ from \citep{mak_ductility_2021} who calibrated $D_\rmc$ so that $D$ predicts Mo and W, to be brittle, and Nb, Ta, and V, to be ductile, in agreement with real experiments at room temperature.
This definition of $D_\rmc$ implicitly introduces the temperature dependence into our model which otherwise depends on the material properties at 0\,K.
Of course, this assumes that $D$ does not change significantly with temperature up to some constant, but, given the good qualitative correlation of $D$ with fracture strains for various alloys at room temperature \citep{hu_screening_2021,mak_ductility_2021}, the choice seems justified to us up to $\sim$\,300\,K.

\begin{figure}[h!]
\centering
\includegraphics[width=0.6\textwidth]{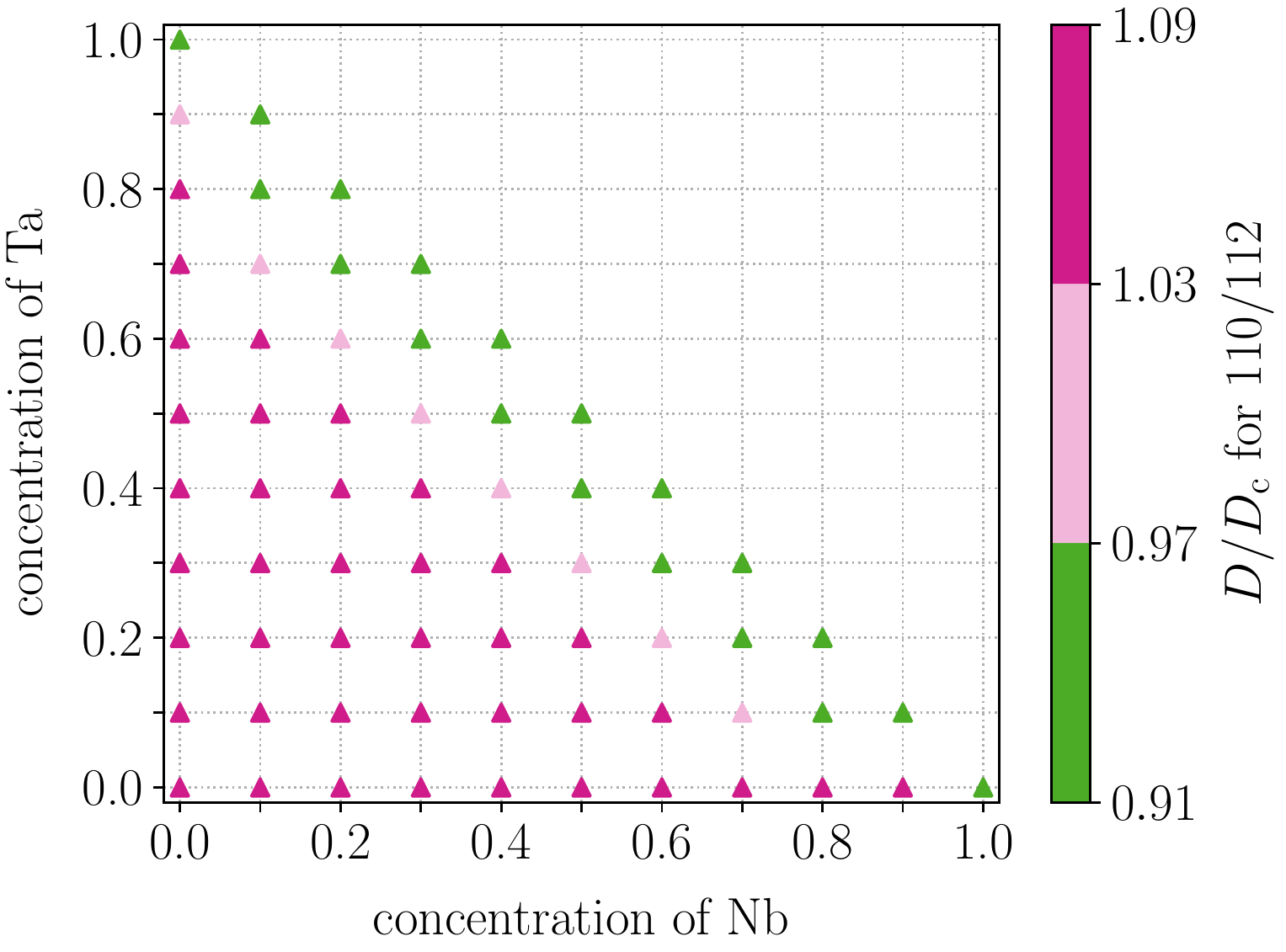}
\caption{Dependence of the ductility on the composition of the Mo-Nb-Ta alloy. All the compositions containing less than 20 \% of Mo, and both Nb and Ta, are ductile.}
\label{Fig:ternary_110_112}
\end{figure}

In Figure \ref{Fig:ternary_110_112}, we show the normalized ductility indices $\bar{D}$\,=\,$D/D_\rmc$ over the entire composition range of Mo-Nb-Ta in 10\,\% intervals.
Therein, we have split the range of the normalized ductility index into three intervals, as done in \citep{mak_ductility_2021}: the interval $0.91\le\bar{D}<0.97$, where the alloy is assumed to be ductile, the interval $0.97\le\bar{D}<1.03$ for borderline ductility, and the interval $1.03\le\bar{D}<1.09$, where the alloy is assumed to be brittle.

With the previous definitions, we conclude the following.
Alloys with high Mo concentration are brittle, as expected, due to the brittleness of Mo.
Vice versa, alloys composed of only Nb and Ta are ductile, since pure-Nb and pure-Ta themselves are ductile.
The most interesting alloys appear to be those with 10\% and 20\% Mo concentration containing both Nb and Ta.
This is due to the fact that, while Mo$_{0.1}$Nb$_{0.9}$ is predicted to be brittle, Mo$_{0.1}$Nb$_{0.8}$Ta$_{0.1}$ is predicted to be ductile.
Moreover, there is a range of compositions Mo$_{0.2}$Nb$_{x}$Ta$_{1-x}$ that are predicted to be borderline ductile.
Hence, some of these alloys could be ductile in practice, while showing higher strength than those from the Mo$_{0.1}$Nb$_{x}$Ta$_{1-x}$ family due to the higher Mo concentration (cf., \citep{statham_thermally_1972}).

From Figures \ref{Fig:Surf} and \ref{Fig:SFE}, we conclude that ductility is more strongly influenced by dislocation emission.
This is because the stacking fault energy decreases roughly twice faster than the surface energy as a function of the Nb-Ta concentration (compare the two upper left plots in Figures \ref{Fig:Surf} and \ref{Fig:SFE}).
Along these lines, we thus remark here that we likely overestimate $D$ with our ductility model based on \emph{average} material properties.
In a random alloy, the stacking fault energy along the crack front is composition-dependent and, thus, local variations in the concentration may favor dislocation nucleation below the average $K_{\rm Ie}$ considered here.
Considering $K_{\rm Ie}$ as a stochastic variable thus appears to us a useful direction for future research to refine the predicted ductility trends.

\section{Concluding Remarks}

In this paper we have presented a high-throughput methodology for predicting the ductility of random alloys and applied it to the Mo-Nb-Ta medium-entropy alloy. The ductility model we employed here is based on the ratio of the $K$-factors for dislocation emission and cleavage---the ductility index---and depends on the stacking fault energies, the surface energies, and the elastic constants. To compute these properties efficiently and with near-DFT accuracy, we used a machine-learning interatomic potential, the Moment Tensor Potential (MTP) \cite{shapeev_moment_2016,gubaev2018-chemoinformatics} as an interatomic interaction model. For constructing the training sets, used to parameterize the MTPs, we used active learning \cite{podryabinkin_active_2017,gubaev2018-chemoinformatics,gubaev_accelerating_2019} to automatically detect those compositions of the random alloy that should be added to the training set. To validate our methodology, the fitted potentials were then used to calculate the stacking fault energies, the surface energies, and the elastic constants, over the entire composition space of Mo-Nb-Ta, and compared with reference (DFT) values from the literature.

Overall, we found that the energies, forces, and stresses, predicted with each fitted MTP are in a good agreement with DFT. We also demonstrated that the stacking fault energies and the surface energies, obtained with each corresponding MTP are close to those predicted with a recently developed surrogate model \citep{hu_screening_2021} that has been fitted to an extensive set of DFT calculations. At the same time, we found that the popular EAM potential of \citet{zhou_misfit-energy-increasing_2004} typically heavily underestimates both the stacking fault energies and the surface energies. On the other hand, the EAM potential predicts the experimental elastic constants, contrary to DFT which underestimates C$_{44}$ \cite{kovci2008-elasticity}, in particular for Nb by more than 100\%. Therefore, while EAM and MTP both predict nearly the same C$_{11}$'s and C$_{12}$'s, MTP constantly underestimates C$_{44}$ over the entire composition space. Thus, we decided to compute the ductility indices using the stacking fault energies and the surface energies calculated with the MTPs, and the elastic constants calculated with the EAM potential. Our main finding here is that the alloys with $\ge$30\% Mo concentration are brittle, but we discovered several potentially ductile or borderline ductile alloys with 10--20\% Mo concentration which contain \emph{both} Nb and Ta.
Since Mo increases the strength, it is in particular those compositions with higher Mo concentration which appear to be the most interesting candidates for novel alloys, possibly possessing excellent strength-vs-ductility ratios.

In addition, we introduced and validated an average-atom MTP, an "alchemical MTP", by averaging the descriptors of the random alloy MTP. This average-atom MTP is a single-component MTP which has been shown to possess nearly the same average material properties as the random alloy, and can be used to further speed up the computations when averaging over random configurations becomes too costly.

Thus, in this paper we demonstrated that our methodology could be considered as a promising tool for screening the brittle/ductile random alloys, although we remark that we had to resort to an EAM potential to reliably compute the elastic constants, because DFT did not predict those.
The problem of DFT not predicting $C_{44}$ is well-known and persists throughout the literature for Nb and V using various established DFT codes \citep{liu_first_2011,de_jong_charting_2015,gao_tests_2020,liao_modeling_2020}.
Hence, should $C_{44}$ be required by a model, we currently recommend using our method only for alloys with no or little Nb-V content.
At this point, we yet remark that the EAM potential of Zhou et al. already covers refractory alloys of Mo-Nb-Ta-V-W.
That is, models depending on the elastic constants for those alloys do not crucially require DFT calculations anymore.
However, we emphasize that this is not a deficiency of our approach as such, but an inaccuracy already present in the DFT model.
From our validation we anticipate that---as long as we are able to fit the MTPs with respect to accurate ab initio calculations---our method can likewise be applied to other interesting alloy combinations, including more than three elements.
More broadly speaking, we further anticipate that our methodology can also be applied to predict other material properties, for instance misfit volumes, to parameterize simple strengthening models \citep{varvenne_theory_2016}, or grain boundary energies, in order to parameterize grain boundary segregation models (e.g., \citep{hu_data-driven_2022}), by adding the relevant structures to the set of training candidates.
In addition, we anticipate that our methodology of training MTPs will prove useful to predict material properties at finite temperature by replacing the relaxation in \textbf{Step\;4} with molecular dynamics simulations.
This will allow for a more accurate parameterization of, e.g., the ductility model considered here, taking temperature effects into account.

Another unique feature of our method is the opportunity to include configurations containing full dislocations in the training using our ``in operando active learning'' technique \citep{hodapp_operando_2020}.
The corresponding MTPs can then be used to predict important properties related to the material's strength, for instance, interactions of dislocations with other types of defects.
In this respect, the average-atom MTP developed here appears to us a promising first step towards quantitatively parameterizing strengthening models \`{a} la \citet{varvenne_theory_2016} which require average values for material properties which are difficult to compute by averaging over random configurations, for example, the average dislocation line tension, or the average change in the dislocation core energy when adding a solute atom to the material.

\section{Materials and Methods}

\subsection{Linear Elastic Fracture Mechanics}
\label{sec:linear_elastic_fracture_mechanics}

\begin{figure}[t]
    \centering
    \includegraphics[width=0.5\textwidth]{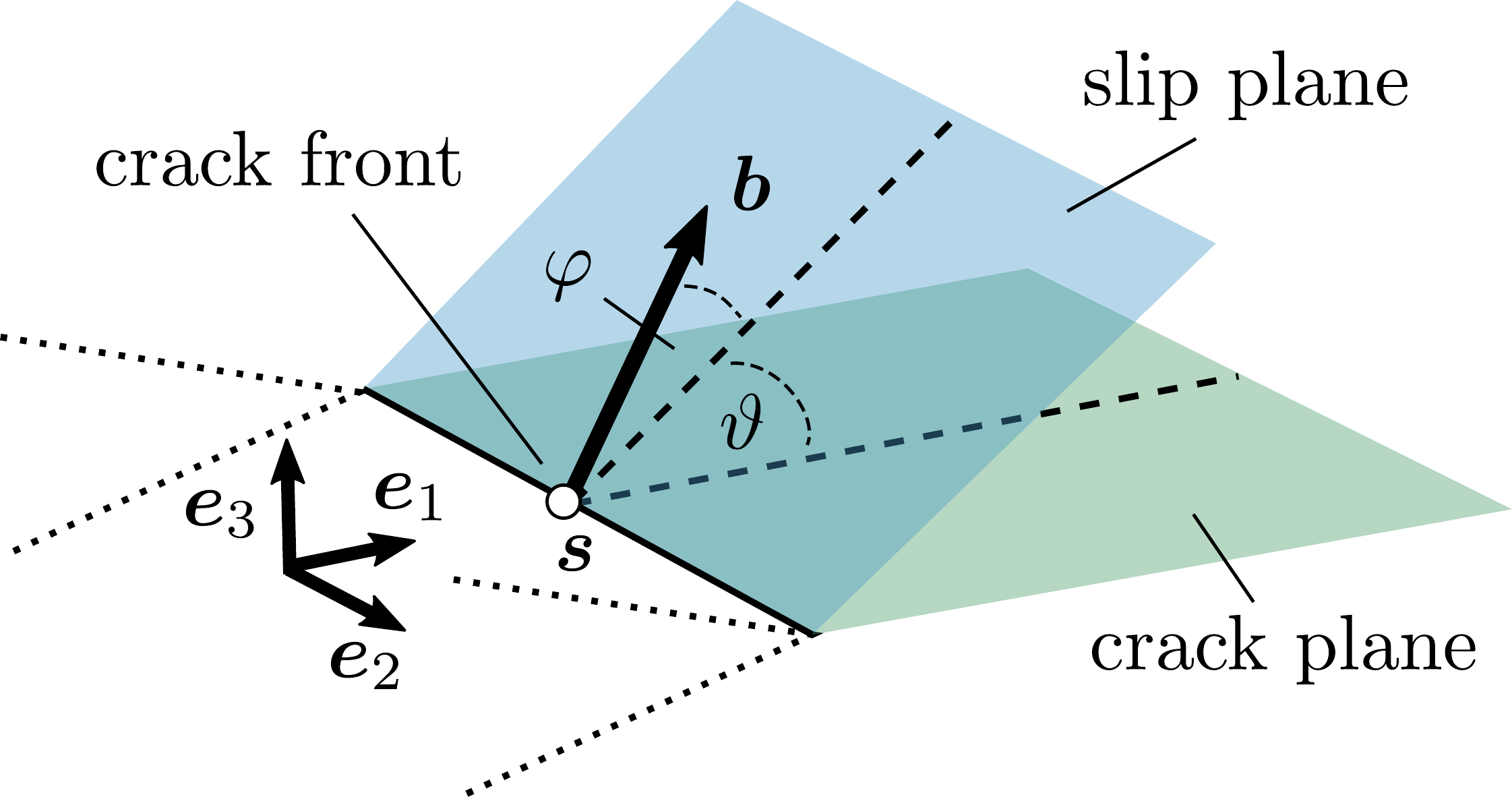}
    \caption{Schematic illustration of a semi-infinite crack}
    \label{fig:crack_front}
\end{figure}

In what follows, we assume a semi-infinite crack with its front located at position $\bms$, as shown in Figure \ref{fig:crack_front}.
We focus on mode I loading.
We assume that the body behaves linear elastic such that the total energy is given by
\begin{equation}
    \Pi = \Pi(\bms, K_{\rm I}) = \frac{1}{2} \int \bmsigma \cdot \bmeps \,\rmd V - \int \bmt \cdot \bmu \,\rmd A,
\end{equation}
with the Cauchy stress $\bmsigma = \bbC[\bmeps]$, the strain tensor $\bmeps$, the tensor of elastic constants $\bbC$, and the surface tractions and displacements $\bmt$ and $\bmu$, respectively.
The total energy depends on the position of the crack, and the external loading, the $K$-factor $K_{\rm I}$.

We are interested in the variation of $\Pi$ due to a change in the position of the crack front along the $\rmx_1$-axis (cf. Figure \ref{fig:crack_front}).
This variation can be expressed in terms of the $J$-integral of \citet{rice_dislocation_1992}
\begin{equation}
    J = \delta_{s_1}\Pi = \underset{\delta s_1 \rightarrow \infty}{\lim} \int_{\Gamma(\delta s_1)} \left( \frac{1}{2} (\bmsigma \cdot \bmeps) n_1 + \bmt \cdot \frac{\partial \bmu}{\partial x_1} \right) \,\rmd A,
\end{equation}
where $\Gamma(\delta s_1)$ is a contour around the crack.
More precisely, $J$ is the energy release per area while the crack propagates.
If the crack propagates along the $\rmx_1$-axis, then
\begin{equation}\label{eq:J_cleavage}
    J = K_{\rm I}^2 \lambda_{22}(\bbC),
\end{equation}
where $\lambda_{22}(\bbC)$ is some quantity depending on the elastic constants.
For dislocation motion, the crack moves along a plane inclined to the crack front.
In this case, $J$ cannot be obtained analytically, so we use the common approximation of \citet{rice_dislocation_1992}
\begin{equation}\label{eq:J_disloc}
    J = \frac{K_{\rm I}^2 F_{12}(\bbC,\mtheta)^2 \cos^2{\phi}}{o(\bbC,\theta,\phi)},
\end{equation}
where $\theta$ is the angle of the slip plane with respect to the crack front, $\phi$ is the angle of the Burgers vector $\bmb$ inclined to the slip direction (see Figure \ref{fig:crack_front}), and $F_{12}$ and $o$ are functions depending on $\theta$, $\phi$, and $\bbC$.
For a precise derivation of \eqref{eq:J_cleavage} and \eqref{eq:J_disloc}, the reader is referred to the textbooks of \citet{ting_anisotropic_1996}, \citet{sun_fracture_2012}, or the review article of \citet{andric_atomistic_2019}.
In particular, the derivation of the functions $\lambda_{22}(\bbC)$, $F_{12}(\bbC,\mtheta)$, and $o(\bbC,\theta,\phi)$, can be found in Section 2.5 in \citep{andric_atomistic_2019}.

For cleavage, $J = 2\gamma_{\rm surf}$, where $\gamma_{\rm surf}$ is the surface energy.
In case of dislocation nucleation, $J = \gamma_{\rm sf}$, where $\gamma_{\rm sf}$ is the stacking fault energy.
Thus, the $K$-factors for cleavage fracture and dislocation emission are given by (e.g., \citep{andric_atomistic_2019})
\begin{align}\label{eq:K_Ic_and_K_Ie}
    K_{\rm Ic} = \sqrt{\frac{2\gamma_{\rm surf}}{\lambda_{22}(\bbC)}},
    &&
    K_{\rm Ie} = \frac{\sqrt{\gamma_{\rm sf} o(\bbC,\theta,\phi)}}{F_{12}(\bbC,\theta) \cos{\phi}}.
\end{align}

With \eqref{eq:K_Ic_and_K_Ie}, we then define the ductility index analogously to \citep{mak_ductility_2021} as
\begin{equation}\label{eq:d_index}
    D = \frac{K_{\rm Ie}}{K_{\rm Ic}} = \chi(\bbC,\theta,\phi) \sqrt{\frac{\gamma_{\rm sf}}{\gamma_{\rm surf}}},
\end{equation}
where
\begin{equation}
    \chi(\bbC,\theta,\phi) = \frac{\sqrt{o(\bbC,\theta,\phi)\lambda_{22}(\bbC)}}{\sqrt{2} F_{12}(\bbC,\theta) \cos{\phi}}
\end{equation}
is some pre-factor depending on the elastic constants and the crack geometry.

\subsection{Atomistic Inputs}

In the following we briefly describe our setup for computing $\gamma_{\rm sf}$, $\gamma_{\rm surf}$, and $\bbC$.
To that end, we assume a configuration $\Atoms$ composed of atoms $\Atom_i$.
The configuration of atoms is further assumed to reside in a supercell and subject to periodic boundary conditions.
Moreover, every configuration has a total energy, depending on $\Atoms$, that we denote by $\Pi(\Atoms)$.

\begin{figure}[hbt]
    \centering
    \includegraphics[width=0.9\textwidth]{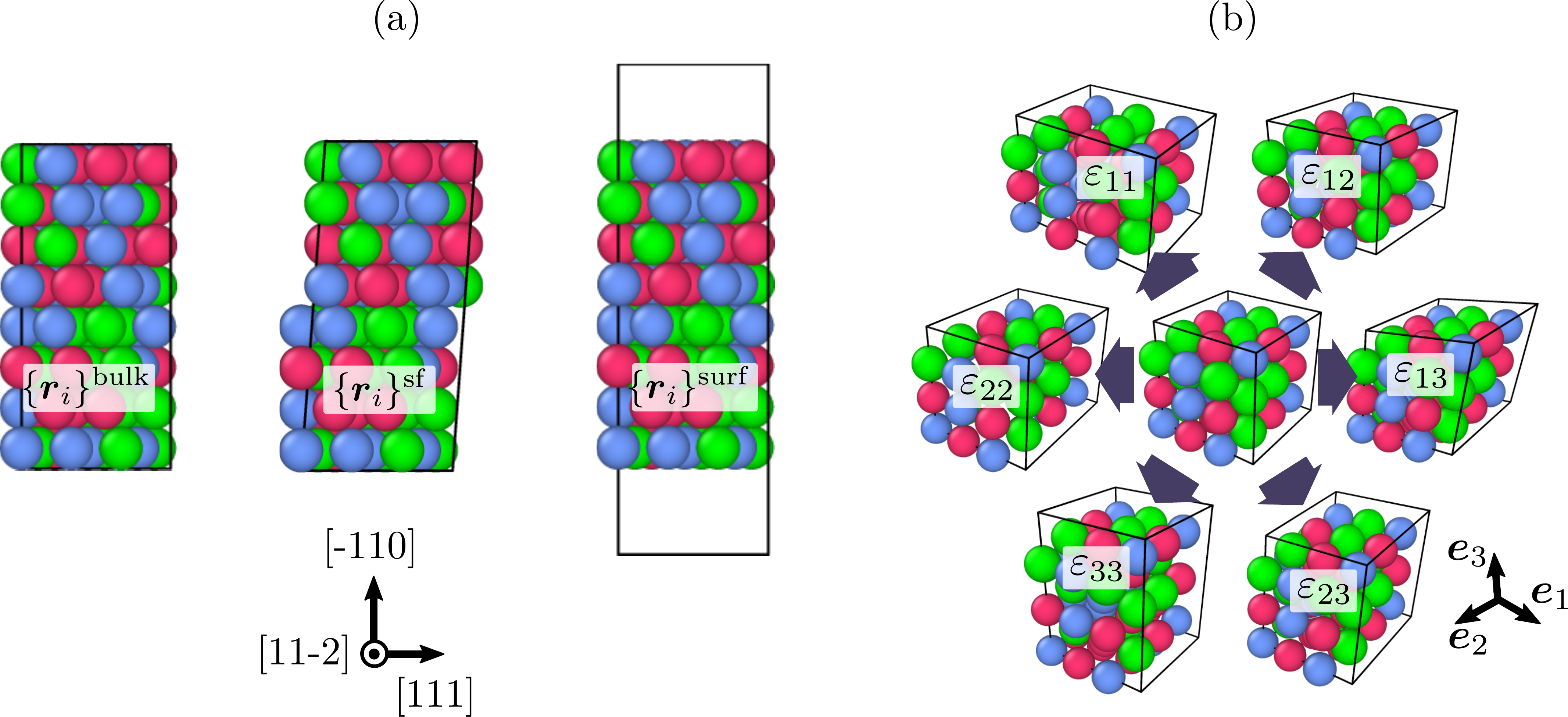}
    \caption{(a)\;Supercells for computing the bulk, stacking fault, and surface energies, exemplified here for configurations with \{110\} being the stacking fault and surface plane, respectively.\;(b)\;Strained configurations for computing the elastic constants}
    \label{fig:cfgs}
\end{figure}

In the following we assume a body-centered-cubic (bcc) lattice.
In order to compute surface and stacking fault energies, we first generate a rectangular bulk configuration, with the $\bme_3$-axis corresponding to the stacking fault or surface plane, respectively, as shown in Figure \ref{fig:cfgs} (a).
We then relax the atoms and denote the relaxed configuration by $\Atoms^{\rm bulk}$.

To compute the stacking fault energy, we translate half of the bulk crystal by half of the Burgers vector.
We further shear the supercell to ensure a perfect crystalline environment at the upper and lower cell boundaries.
This procedure generates a periodic configuration with one single stacking fault in the center of the cell.
We, again, relax the atoms in the direction normal to the slip plane and denote the relaxed configuration by $\Atoms^{\rm sf}$.
We then compute the stacking fault energies as follows
\begin{equation}
    \gamma_{\rm sf} = \frac{\Pi(\Atoms^{\rm sf}) - \Pi(\Atoms^{\rm bulk})}{A},
\end{equation}
where $A$ is the area of the slip plane.

To compute the surface energy, we extend the supercell with the bulk configuration in the $\bme_3$-direction.
We then relax the atoms and denote the relaxed configuration again by $\Atoms^{\rm bulk}$.
The surface energy is then given by
\begin{equation}
    \gamma_{\rm surf} = \frac{\Pi(\Atoms^{\rm surf}) - \Pi(\Atoms^{\rm bulk})}{2A},
\end{equation}
where $A$ now corresponds to the area of the surface plane.

Finally, to compute the elastic constants, we apply six linear independent infinitesimal shear strains to a bulk configuration (Figure \ref{fig:cfgs} (b)).
We then compute the stresses for each sheared configuration, and collect the six strains and stresses in the matrices $\mathsf{E}$ and $\mathsf{S}$, respectively.
We then compute the $6\times 6$ matrix of elastic constants, corresponding to $\bbC$, as $\mathsf{C}=\mathsf{S}\mathsf{E}^{-1}$.

We remark that, to compute average material properties of a random alloy, we typically generate a set of different configurations with the elements randomly distributed to the atoms according to the desired composition, and average over the obtained $\gamma_{\rm sf}$'s, $\gamma_{\rm surf}$'s, and $\bbC$'s.

For the relaxations, we have used the Fast Inertial Relaxation Engine \citep[FIRE,][]{bitzek_structural_2006}, as implemented within the ASE package \citep{hjorth_larsen_atomic_2017}.
We usually terminated the relaxations when the maximum force on all atoms was smaller than 0.001\,eV/\AA.
For computing the elastic constants, we have used the \texttt{get\_elastic\_constants} function from Matscipy (\url{https://github.com/libAtoms/matscipy}).

\subsection{Moment Tensor Potentials}

For accelerating expensive DFT calculations, and calculating the ductility indices, we use the Moment Tensor Potential (MTP) \cite{shapeev_moment_2016, gubaev2018-chemoinformatics} that we fit to DFT data.  The MTP is a local potential as its total energy is a sum of contributions of each $i$-th atom:

\begin{equation} \label{MTP}
    \Pi^{\rm MTP} = \sum \limits_i \sum \limits_{\alpha} \xi_{\alpha} B_{\alpha} ({\bf \mathfrak{n}}_i),
\end{equation}
where $\xi_{\alpha}$ are the (linear) MTP parameters, and $B_{\alpha}({\bf \mathfrak{n}}_i)$ are the basis functions that depend on the neighborhood ${\bf \mathfrak{n}}_i$ of each $i$-th atom. The neighborhood includes all the $j$-th atoms for which the distance $|{\bf r}_{ij}|$ between the $i$-th and $j$-th atom is smaller than the radius $R_{\rm cut}$. We construct the scalar basis functions by performing contractions of the Moment Tensor Descriptors:

\begin{equation} \label{Descriptor}
    M_{\mu, \nu}({\bf \mathfrak{n}}_i) = \sum \limits_j f_{\mu}(|{\bf r}_{ij}|) {\bf r}_{ij}^{\otimes \nu}.
\end{equation}
The descriptors contain the radial and the angular part. The radial part describes the two-body interactions and has the form:

\begin{equation} \label{RadialPart}
    f_{\mu}(|{\bf r}_{ij}|) = \sum \limits_{\beta} c_{\mu, t_i, t_j}^{\beta} T^{\beta}(|{\bf r}_{ij}|) (R_{\rm cut} - |{\bf r}_{ij}|)^2,
\end{equation}
where $\mu$ is the number of the radial function, $t_i$ and $t_j$ are atomic types, $T^{\beta}$ is the Chebyshev polynomial of $\beta$-th order, and $c_{\mu, t_i, t_j}^{\beta}$ are the (radial) MTP parameters. The radial part and, therefore, the Moment Tensor Descriptor tends to zero when $|{\bf r}_{ij}|$ tends to $R_{\rm cut}$. The angular part ${\bf r}_{ij}^{\otimes \nu}$ describes many-body interactions and by definition:

\begin{equation} \label{AngularPart}
    {\bf r}_{ij}^{\otimes \nu} = \underbrace{{\bf r}_{ij} \otimes \ldots \otimes {\bf r}_{ij}}_{\nu},
\end{equation}
where ``$\otimes$'' is the outer product of vectors. The numbers $\mu$ and $\nu$ determine the so-called level of the Moment Tensor Descriptor ${\rm lev} M_{\mu, \nu} = 2 + 4 \mu + \nu$ and the level of the MTP basis function ${\rm lev} B_{\alpha} = \sum \limits_{m=1}^{N_M} (2 + 4 \mu_m + \nu_m)$, where $N_M$ is the number of Moment Tensor Descriptors included in the contraction yielding a scalar, the MTP basis functions. To construct a particular functional form for the MTP we choose the MTP level, ${\rm lev}_{\rm MTP}$, and include in \eqref{MTP} only the basis functions with ${\rm lev} B_{\alpha} \leq {\rm lev}_{\rm MTP}$.

We combine the linear parameters $\xi_{\alpha}$ and the radial parameters $c_{\mu, t_i, t_j}^{\beta}$ in a set ${\bm \mtheta}$ containing all free parameters and denote the MTP energy by $\Pi^{\rm MTP} = \Pi({\bm \mtheta})$. The free parameters are determined by solving the optimization problem:

\begin{equation} \label{Fitting}
\displaystyle
\sum \limits_k \Bigl[
	w_{\rm e} \left(\Pi_k ({\bm {\mtheta}}) - \Pi^{\rm DFT}_k \right)^2
	+
	w_{\rm f} \sum_{i} \left| {\bm f}_{k,i}({\bm {\mtheta}}) - {\bm f}^{\rm DFT}_{k,i} \right|^2 
	+
	w_{\rm s} \big|\bmsigma_k(\bm {\mtheta}) - \bmsigma_k^{{\rm DFT}}\big|^2 \Bigr] \to \min\limits_{\bm \mtheta},
\end{equation}
where $k$ is the number of a configuration in the training set, $\Pi^{\rm DFT}_k$, ${\bm f}^{\rm DFT}_{k,i}$, and $\sigma_k^{{\rm DFT}}$ are the energies, forces, and stresses calculated with DFT, and $w_{\rm e}$, $w_{\rm f}$, and $w_{\rm s}$ are the non-negative weights which express the importance of fitting MTP energies, forces, and stresses to their DFT counterparts.

\subsection{Active Learning}

For constructing the training set we use the active learning algorithm based on the D-optimality criterion from \cite{podryabinkin_active_2017, gubaev2018-chemoinformatics}. This algorithm allows us to select the most geometrically different (or, representative) configurations and, thus, to reduce the number of expensive DFT calculations in comparison with passive learning where we manually construct the training set.

Assume we have an initial training set of $K$ (unrelaxed) configurations, an MTP with $m$ parameters fitted to this training set, and $K \geq m$. We start from constructing the so-called active set using the MaxVol algorithm \cite{goreinov2010-maxvol}. The active set contains a subset of $m$ configurations from the training set which maximizes the volume of the matrix $\mathsf{A}$, or, det$|\mathsf{A}|$:

\[
\mathsf{A}=\left(\begin{matrix}
\frac{\partial \Pi_1}{\partial \mtheta_1}\left( {\bm \mtheta} \right) & \ldots & \frac{\partial \Pi_1}{\partial \mtheta_m}\left({\bm \mtheta} \right) \\
\vdots & \ddots & \vdots \\
\frac{\partial \Pi_m}{\partial \mtheta_1}\left({\bm \mtheta} \right) & \ldots & \frac{\partial \Pi_m}{\partial \mtheta_m}\left({\bm \mtheta} \right) \\
\end{matrix}\right),
\]
where each row of the matrix $\mathsf{A}$ corresponds to a particular configuration. We note that in the case $K < m$ we fill in the first lines of the matrix $\mathsf{A}$ with the configurations that maximize the volume of the matrix $\mathsf{A}$. We fill the rest of the lines with ones on the diagonal and zeros outside.

Next, we create a big set of the candidate configurations. The number of these configurations is much larger than the number of configurations in the initial training set. From this set with candidate configurations we select the most representative configurations, and add them to the training set. To that end, we introduce the extrapolation grade:
\begin{equation} \label{Grade}
\begin{array}{c}
\displaystyle
\gamma = \max_{1 \leq j \leq m} (|c_j|), ~c = \left( \dfrac{\partial \Pi^*}{\partial \mtheta_1} (\bm \mtheta) \ldots \dfrac{\partial \Pi^*}{\partial \mtheta_m} (\bm \mtheta) \right) \mathsf{A}^{-1}.
\end{array}
\end{equation}
Here $\Pi^*(\bm \mtheta)$ is the energy of the candidate configuration predicted with MTP, i.e., we do not need to conduct DFT calculations for estimating $\gamma$. The extrapolation grade \eqref{Grade} describes how much this configuration is geometrically different from the configurations in the active set, and this difference increases with increasing grade.

We, further, iteratively select and add the configurations to the training set starting from the candidates with the highest extrapolation grades, i.e., we first add only the configurations with $\gamma > 10000$, next the ones with $\gamma > 1000, 100, 10, 1$. We do not consider those configurations with $\gamma < 1$, since any configuration with such a small extrapolation grade does not geometrically differ from the ones in the active set. Thus, we have an updated training set that is expected to cover \emph{almost the whole configuration space} of interest. We fit the MTP to the data of the updated set, and re-initialize the active set.

Finally, we relax all the configurations in the updated training set with respect to the atomic positions and actively select the remaining extrapolative configurations to be added to the training set. The MTP, trained on this training set is then expected to cover the \emph{whole configuration space}.

\subsection{Average-atom MTPs}

There are material properties where averaging over configurations can become very costly, or the averaging procedure itself can become too cumbersome.
One example is the average line tension of a random alloy \citep{varvenne_theory_2016}.
For such cases, it can be convenient to average already on the potential level.
A potential-averaging procedure was introduced  by \citet{smith_application_1989} for EAM potentials, and has recently been proven valuable for parameterizing strengthening models for random alloys \citep{varvenne_theory_2016}.

However, such a procedure is not necessarily limited to EAM potentials.
The easiest way to construct an average-atom MTP is to average over all possible neighborhoods for a given composition, that is,
\begin{equation}
 \langle M_{\mu, \nu}({\bf \mathfrak{n}}_i) \rangle
 =
 \sum_{X,Y} \mathfrak{c}_X \mathfrak{c}_X M_{\mu, \nu}({\bf \mathfrak{n}}_i)
 =
 \sum_{X,Y} \mathfrak{c}_X \mathfrak{c}_Y \sum \limits_{j,\beta} c_{\mu, X, Y}^{\beta} T^{\beta}(|{\bf r}_{ij}|) (R_{\rm cut} - |{\bf r}_{ij}|)^2 {\bf r}_{ij}^{\otimes \nu},
\end{equation}
where $\mathfrak{c}_X$ and $\mathfrak{c}_Y$ are the concentrations of an element $X$ and $Y$, respectively.
The descriptors of the average-atom MTP are then the averaged $M_{\mu, \nu}$'s.
This is convenient because the parameters $\bar{c}_\mu^{\beta}$ of this new (averaged) MTP can be pre-computed as
\begin{equation}
 \bar{c}_\mu^{\beta} = \sum_{X,Y} \mathfrak{c}_X\mathfrak{c}_Y c_{\mu, X, Y}^{\beta}
\end{equation}
such that the average-atom MTP can be used just as any other single-component MTP.

The underlying assumptions for this procedure to work are (i) that the influence of local relaxations of the random alloy remains sufficiently small, and (ii) that the descriptors are uncorrelated.
For example, for a basis function that is constructed by contracting two descriptors, $M_1$ and $M_2$, we need to assume that their covariance
\begin{equation}
 \operatorname{cov}(M_1,M_2) = \langle M_1 \cdot M_2 \rangle - \langle M_1 \rangle \langle M_2 \rangle
\end{equation}
is sufficiently small.
It is easy to see that the averaging is exact for pair potentials (since there is only one descriptor), but not for the general case of many-body interactions.
The numerical experiments, presented in this work, however, show that the descriptors indeed appear to be uncorrelated.

To make the averaging procedure more robust, one possibility is to subject the parameter fitting \eqref{Fitting} to a minimization of $\operatorname{cov}(M_1,M_2)$.
Since the parameter set is not unique (there are many practically equally good parameter sets for a given training set), it is possible that such a constrained fitting could yield a good average-atom MTP.
We leave this for exploration in future work.

\subsection{VASP Calculations}

We conducted VASP calculations with the PBE functional and the PAW pseudopotentials. The energy cut-offs were 505 eV, 470 eV, and 503 eV for Mo, Nb, and Ta, respectively. The $k$-point spacing was of 0.15 $\angstrom^{-1}$. Self-consistent atomic relaxation was terminated when the energy difference between
two subsequent iterations was less than $10^{-4}$ eV.

\section{Acknowledgements}

This work was supported by the Russian Science Foundation (grant number 18-13-00479).
MH would further like to thank W. A. Curtin and B. Yin for helpful discussions during the initial stage of this project.
Moreover, the authors would like to thank P. Andric for providing his scripts for computing the Stroh tensors for the $K$-factors.

\section{Data Availability}

The implementation of the MTPs is included in the MLIP package which is publicly available for academic use at \url{https://mlip.skoltech.ru/download/} upon registration. Additional scripts, necessary to run our algorithm, as well as the training data, are available from the authors upon reasonable request.

\nocite{stukowski_visualization_2010}

\bibliographystyle{apalike}
\bibliography{mtp-hea-ductility}

\end{document}